M. Arminjon, D. Imbault

# Maximum entropy principle and texture formation

*The macro-to-micro transition in a heterogeneous material is envisaged as the selection of a probability distribution by the Principle of Maximum Entropy (MAXENT). The material is made of constituents, e.g. given crystal orientations. Each constituent is itself made of a large number of elementary constituents. The relevant probability is the volume fraction of the elementary constituents that belong to a given constituent and undergo a given stimulus. Assuming only obvious constraints in MAXENT means describing a maximally disordered material. This is proved to have the same average stimulus in each constituent. By adding a constraint in MAXENT, a new model, potentially interesting e.g. for texture prediction, is obtained.*

## 1    The principle of maximum statistical entropy (MAXENT)

MAXENT gives a link between information theory and statistical mechanics [1]. According to *information theory,* the "amount of uncertainty" represented by a probability distribution $(p_i)_{i=1,...,M}$ on a finite set $E = \{x_1, ..., x_M\}$ is the *statistical entropy* given by

$$S = -\sum_{i=1}^{M} p_i \operatorname{Log} p_i. \tag{1}$$

MAXENT applies to the case where only some expectation values

$$\langle \phi_q \rangle \equiv \sum_{i=1}^{M} p_i \, \phi_q(x_i) = a_q \quad (q = 1, ..., Q) \tag{2}$$

are known (with $\phi_q$ known functions and $Q \ll M$). In a such case, it is clear that the distribution $(p_i)_{i=1,...,M}$ is not determined by the data $a_q$ ($q = 1, ..., Q$). MAXENT says that *the relevant distribution $(p_i)$ makes $S$ a maximum with the $Q$ constraints (2).* This amounts to selecting the broadest probability distribution compatible with the available information. *Statistical mechanics,* on the other hand, considers a system made of a huge number $N$ of "elementary constituents", *e.g.* molecules (in the kinetic theory of gases). The *micro-state* (velocity and position) of each molecule is in one among $M$ possible boxes, with $1 \ll M \ll N$. Let $l_i$ ($i = 1, ..., M$) be the number of molecules in box ($i$). The corresponding fraction is $p_i = l_i/N$ (thus $p_i \geq 0$ and $p_1 + ... + p_M = 1$, as required for a probability distribution). The *macro-state* is a set of relevant macroscopic parameters: pressure, density, temperature, ... Each parameter making the macro-state should be computable from the probability distribution $(p_i)$, as the average (expectation) $\langle \phi \rangle$ of some known function $\phi$. Now a given probability distribution $(p_i)$ is obtainable by a large number of distinct *configurations* [a configuration is the mapping: molecule $\rightarrow$ box ($i$)]. The hypothese made in statistical mechanics is that the "real" distribution is the one that may be obtained by the largest number of distinct configurations. Since the possible configurations must be compatible with the given macro-state, the most general version of this hypothese is MAXENT [1-2]. Indeed MAXENT has become the most fundamental principle in statistical physics [2].

## 2    Implementation of MAXENT in physics of heterogeneous media

Consider a heterogeneous medium, *e.g.* a *polycrystal*, with two microscopic fields: stimulus and response, *e.g.* strain-rate **d** and *stress* **σ**. (In a porous medium, these would be replaced by the pressure gradient and the filtration velocity. It would be easy to give many more examples.) The ideal goal of the *macro-to-micro* transition is to determine the micro-fields **d(x)**, **σ(x)** from the mere data of the macro-stimulus, say **D**. This seems to be an unattainable goal, the more so if the microscopic constitutive relation is non-linear [3]. In order that the macro-to-micro transition in the polycrystal (for example) may fit with the MAXENT procedure, the *first step* is to *define the micro-state* in the heterogeneous medium. This we define as the joint data of the microscopic stimulus [thus here the value of **d(x)**] and the local state **X(x)** in the heterogeneous medium [3], with **X(x)** being the set of the internal and/or geometrical variables that make the microscopic constitutive law depend on the micro position **x** [4]. For a polycrystal, the heterogeneity is mainly due to the anisotropy of the constitutive crystals. The microscopic constitutive law **d(x)** – **σ(x)** is hence often considered to be known from the mere data of the local orientation **R(x)**, which means that **X(x)** = **R(x)**. We shall use this assumption to fix the ideas in the following, but it is by no means necessary to the general method discussed. Thus in a deformed polycrystal, the micro-state is ($\mathbf{d}(\mathbf{x})$, $\mathbf{R}(\mathbf{x})$). The *second step* is to *discretize the possible values of the micro-state*. First, we assume a discrete orientation distribution:

at time $t$, $\mathbf{R}(\mathbf{x}, t) \in \{\mathbf{R}_1(t), ..., \mathbf{R}_n(t)\}$. (3)

[Note that, in an ideal polycrystal, the crystal orientations would have *a priori* a discrete distribution: in that case, the discretization of $\mathbf{R}(\mathbf{x}, t)$ would be trivial. Many procedures exist to discretize the orientation distribution in a real polycrystal, *i.e.*, to approximate the real orientation field by a piecewise constant field satisfying (3).] The volume fractions $f_k$ ($k = 1, ..., n$) of the different orientations are given, with $f_1 + ... + f_n = 1$. The current *texture* may be characterized by the data of ($f_1$, $\mathbf{R}_1(t)$), ..., ($f_n$, $\mathbf{R}_n(t)$). Due to the incompressibility of plastic deformation, the fractions $f_k$ may be assumed constant. Yet the texture evolves due to the evolution of the orientations, *i.e.*, due to the dependence $\mathbf{R}_k = \mathbf{R}_k(t)$.

An attainable aim for the macro-to-micro transition (*e.g.* in a polycrystal), is to calculate the list $(\mathbf{D}_k)_{k=1,...,n}$ with $\mathbf{D}_k = \mathbf{D}_k(t)$ the *average strain-rate in the orientation* $\mathbf{R}_k(t)$. Thus $\mathbf{D}_k(t)$ is the average of the micro-field $\mathbf{d}(\mathbf{x}, t)$ over the zone $Z_k$ of the polycrystal where the crystal orientation is $\mathbf{R}_k(t)$. Then, using the constitutive law for this orientation, the (average) stress $\boldsymbol{\sigma}_k(t)$ is obtained. The (average) rotation rates $\boldsymbol{\Omega}_k(t) = \dot{\mathbf{R}}_k \mathbf{R}_k(t)^{-1}$ are also obtained, hence an evolution (averaged over each orientation) is got. But, in order to use MAXENT so as to calculate the distribution ($\mathbf{D}_k(t)$), we still have to discretize the possible values taken by the strain-rate $\mathbf{d}(\mathbf{x}, t)$. In the following, we consider a fixed time $t$, hence we omit the dependence with $t$ henceforth. Using a (hyper)cubic mesh with a small size $\varepsilon$ for the strain-rate, the discretization is defined by the nodes of the mesh, say $\mathbf{D}^1, ..., \mathbf{D}^m$. We substitute for $\mathbf{d}(\mathbf{x})$ the following piecewise constant field:

$$\mathbf{d'}(\mathbf{x}) = (d'_1(\mathbf{x}), ..., d'_6(\mathbf{x})) \text{ with } d'_l(\mathbf{x}) = \varepsilon \text{ Int } (d_l(\mathbf{x})/\varepsilon) \qquad [(\text{Int}(\xi) = k) \Leftrightarrow (k \text{ integer and } k \leq \xi < k+1)] \quad (4)$$

where $T_1, ..., T_6$ are the independent components of a second-order symmetric tensor $\mathbf{T}$. Thus, the domains

$$\Omega_j \equiv \{\mathbf{x} ; \mathbf{d'}(\mathbf{x}) = \mathbf{D}^j\} \ (j = 1, ..., m) \quad (5)$$

are well-defined, two by two disjoined, and their union covers the whole polycrystal. The same is true for the $Z_k$'s. We have

$$\|\mathbf{d}(\mathbf{x}) - \mathbf{D}^j\|_\infty = \|\mathbf{d}(\mathbf{x}) - \mathbf{d'}(\mathbf{x})\|_\infty \leq \varepsilon \text{ for } \mathbf{x} \in \Omega_j. \quad (6)$$

Here $\|\mathbf{T}\|_\infty \equiv \max(|T_1|, ..., |T_6|)$ for a second-order symmetric tensor $\mathbf{T}$. The relevant probability distribution is defined as

$$p^j{}_k = \text{volume fraction of } \Omega_j \cap Z_k \text{ in the polycrystal} = V(\Omega_j \cap Z_k) / V(\Omega). \quad (7)$$

Thus $p^j{}_k$ is the probability of the joint event $\mathbf{d'}(\mathbf{x}) = \mathbf{D}^j$ and $\mathbf{R}(\mathbf{x}) = \mathbf{R}_k$, the probability being defined as the volume fraction. [*I.e.*, $P(A) \equiv V(A)/V(\Omega)$, $\Omega$ being the considered representative volume element (RVE) of the polycrystal. Strictly speaking, the notion of RVE is an asymptotic one [4-5], but a simpler illustration is got if exactly representative volume elements like $\Omega$ are assumed to exist.] The domains $\omega_{jk} = \Omega_j \cap Z_k$ are the elementary constituents. They depend on the discretization imposed to the strain-rate field, *i.e.*, they depend on the small parameter $\varepsilon$. Hence the micro-state ($\mathbf{d'}$, $\mathbf{R}$) belongs to the finite product set $\{\mathbf{D}^1, ..., \mathbf{D}^m\} \times \{\mathbf{R}_1, ..., \mathbf{R}_n\}$, so $i$ has become ($j, k$) and $M = m \times n$ in eq. (1). The volume average of the strain-rate $\mathbf{d'}$ in orientation $\mathbf{R}_k$ is given (*cf.* Bayes' conditional probability formula) by

$$\mathbf{D}_k = (p^1{}_k \mathbf{D}^1 + ... + p^m{}_k \mathbf{D}^m)/f_k. \quad (8)$$

Now we want to use MAXENT to determine the discrete probability distribution $(p^j{}_k)$. There are two obvious constraints:

i) The volume fraction of polycrystal in the orientation $\mathbf{R}_k$ is the data $f_k = V(Z_k) / V(\Omega)$. Since the domains $\Omega_j$ are two by two disjoined, and since their union covers the whole polycrystal (or the RVE $\Omega$), we get from (7):

$$\sum_{j=1}^{m} p^j{}_k = f_k \qquad (k = 1, ..., n). \quad (9)$$

ii) The average strain-rate is the applied macroscopic strain-rate $\mathbf{D}$:

$$\sum_{j,k} p^j{}_k \mathbf{D}^j = \mathbf{D}. \quad (10)$$

Hence we may define a model based on MAXENT as follows:

*Model (a):* Maximize $S = -\sum_{k=1}^{n} \sum_{j=1}^{m} p^j_k \, \text{Log} \, p^j_k$ under constraints (9) and (10). (11)

Since the statistical entropy $S$ is a measure of disorder, the latter MAXENT model with obvious constraints may be called the "*volume-fraction model with maximum disorder*" [the disorder is that of the strain-rate distribution ($p^j_k$)]. It is often said that the self-consistent models describe a situation with *perfect disorder* (this is the ideal situation where spatial correlations of a finite range do not exist). Indeed there are arguments showing that the classical self-consistent model for linear elasticity may correspond to that ideal situation [6]. Taking words naively, one might then wonder if the volume-fraction model with maximum disorder is something like a self-consistent model. We prove below that it is not the case.

## 3     The volume-fraction model with maximum disorder is "the Voigt-Taylor model plus random fluctuations"

*I.e.,* this model (eq. (11)) leads to the following prediction for the average strain-rate $\mathbf{D}_k$ in orientation $\mathbf{R}_k$, defined by (8):

$\mathbf{D}_k = \mathbf{D}$ for all $k = 1, ..., n.$ (12)

P r o o f. We use the method of Lagrange multipliers to find the maximum (11): any solution of (11) is a stationary point of

$$\Phi \equiv -\sum_{j,k} p^j_k \, \text{Log} \, p^j_k - \sum_{k=1}^{n} \lambda_k \left( f_k - \sum_{j=1}^{m} p^j_k \right) - \sum_{l=1}^{6} \mu_l \left( D_l - \sum_{j,k} p^j_k D^j_l \right) \quad (13)$$

(with $D_l$ the $l^{\text{th}}$ component of $\mathbf{D}$ and $D^j_l$ the $l^{\text{th}}$ component of $\mathbf{D}^j$). That is, we must have $\partial \Phi / \partial p^j_k = 0$. This is equivalent to

$$p^j_k = e^{\lambda_k - 1} \exp\left( \sum_l \mu_l D^j_l \right) = p^j_k(\lambda_k, (\mu_l)). \quad (14)$$

The multipliers $\lambda_k$ and $\mu_l$ are determined by the condition that the $p^j_k$'s making $\Phi$ stationary (eq. (14)) satisfy the constraints (9) and (10). In the present case, inserting (14) into the constraint (9) allows to eliminate $\lambda_k$ by

$$e^{\lambda_k - 1} = f_k \Bigg/ \sum_{j=1}^{m} \exp\left( \sum_{l=1}^{6} \mu_l D^j_l \right). \quad (15)$$

Calculating $D_{k\,l} \equiv (\mathbf{D}_k)_l = \sum_j p^j_k D^j_l / f_k$ using (14) and (15) gives

$$D_{k\,l} = \sum_{j=1}^{m} \frac{D^j_l \exp\left( \sum_{l'=1}^{6} \mu_{l'} D^j_{l'} \right)}{\sum_{j'=1}^{m} \exp\left( \sum_{l'=1}^{6} \mu_{l'} D^{j'}_{l'} \right)}. \quad (16)$$

Since this does not depend on $k$ and since $\sum_k f_k \mathbf{D}_k = \mathbf{D}$ by eqs. (8) and (10), it follows that $\mathbf{D}_k = \mathbf{D}$ for all $k$. Q.E.D..

    Thus, model (a) is analogous to the Voigt-Taylor model *in the sense of eq. (12)*, though it also describes some random fluctuation, in each constituent, of the stimulus field **d** around the macroscopic stimulus **D**.

## 4     A more interesting MAXENT model (one more constraint)

Equation (12) is unrealistic. To obtain a better model, we must add information, *i.e.* add constraints in MAXENT. One possible new constraint is to impose that the average potential is known (assuming therefore that the micro-law expressing $\sigma(\mathbf{x})$ as a function of $\mathbf{d}(\mathbf{x})$ does derive from a potential, say $u_k$ in the orientation $\mathbf{R}_k$) [3] :

$$\langle u \rangle \equiv \sum_k f_k \, u_k(\mathbf{D}_k) \equiv \sum_k f_k \, u_k \left( \sum_j p^j{}_k \, \mathbf{D}^j / f_k \right) \equiv U(\mathbf{D}) \text{ known} \tag{17}$$

This is the minimum information to add in order to determine the macroscopic behavior since, in the most favourable case, the average potential is indeed a potential for the macro-law [5, 7]. Yet it means the *micro-to-macro* transition is solved! In a previously studied "inhomogeneous variational model" (IVM), the data $U(\mathbf{D})$ may be replaced by the data of the average heterogeneity $h$, with (for some real exponent $p \geq 1$, depending on the behavior of the $u_k$ potentials at large $\mathbf{d}$ [4-5])

$$h^p \equiv \sum_{k=1}^n f_k \|\mathbf{D}_k - \mathbf{D}\|^p \tag{18}$$

It has been shown in Ref. [3] that the *macro-to-micro* transition, *i.e.,* determining the distribution $(\mathbf{D}_k)_{k=1,\ldots,n}$ from data $\mathbf{D}$ (plus necessary additional data: either $h$ or $U(\mathbf{D})$, in the present case), is very close in that model and in the above MAXENT model based on constraints (9), (10) and (17). But it is simpler to impose directly the average heterogeneity $h$, because it is computable from the unknown $(p^j{}_k)$ of the MAXENT procedure. Indeed, using eq. (8), one finds easily that

$$h^p = \sum_{k=1}^n \left\| \sum_{j=1}^m p^j{}_k \mathbf{D}^j - f_k \mathbf{D} \right\|^p / f_k{}^{p-1}. \tag{19}$$

Thus we propose a *new model,* that consists in adding the constraint $h = r$ (with $r$ a given number) in model (a) defined by eq. (11). This new model does not need that the micro-law derives from a potential. Actually, in this purely statistical model, the micro-law itself influences the strain distribution very indirectly – through the value $r$ of the actual heterogeneity, which in reality does indeed depend on the micro-law (and on the geometry), but which is considered, in the new model, as *the* relevant information (in addition to the volume fractions). Note that the actual heterogeneity $r$ is measurable. But, to make use of the new model, $r$ should rather be phenomenologically assumed, as is also done in the IVM [4]. To study rotation effects, one may think to substitute the velocity gradient $\mathbf{l}$ (with 9 independent components) for $\mathbf{d}$ [8].

## 5  Conclusions

i) A general formulation of the Maximum Entropy Principle (MAXENT) has been given for the macro-to-micro transition in a heterogeneous medium. This formulation was illustrated for a textured polycrystal with inelastic deformation.

ii) MAXENT demands constraints. The most obvious ones (the volume fractions are imposed, and the macro-average of the micro-stimulus is the macro-stimulus) lead to predict the same average stimulus in each constituent (as in Voigt's model).

iii) Imposing the value of the average potential gives [3] a model close to the inhomogeneous variational model [4-5]. But the new model proposed consists in imposing directly the average heterogeneity as an additional constraint in MAXENT.

iv) MAXENT provides a general method to build more and more accurate models by adding information (*i.e.* constraints).

*Address:* Lab. "Sols, Solides, Structures", Institut de Mécanique de Grenoble, BP 53, 38041 Grenoble cedex 9, France.